\documentclass[11pt,graphicx]{article}
\usepackage{amssymb,amsmath,amsfonts}
\usepackage{graphicx}
\usepackage{graphics}
\usepackage{eepic,epsfig}
\newcommand{\n}{{\not\hspace{-0.5ex}\nabla}}
\newcommand{\A}{{\not\hspace{-0.8ex}A}}

\begin{document}
\title{Vacuum polarization by fermionic fields in higher dimensional cosmic string space-time}
\author{J. Spinelly$^{1}$ {\thanks{E-mail: jspinelly@uepb.edu.br}}  
and E. R. Bezerra de Mello$^{2}$ \thanks{E-mail: emello@fisica.ufpb.br}\\
1.Departamento de F\'{\i}sica-CCT, Universidade Estadual da Para\'{\i}ba\\
Juv\^encio Arruda S/N, C. Grande, PB, Brazil\\
2.Departamento de F\'{\i}sica-CCEN, Universidade Federal da Para\'{\i}ba\\
58.059-970, C. Postal 5.008, J. Pessoa, PB,  Brazil}

\maketitle
\begin{abstract}
In this paper we investigate vacuum polarization effects associated with charged massive quantum fermionic fields in a six-dimensional cosmic string space-times considering the presence of a magnetic flux running along its core. We have shown that for specific values of the parameters which codify the presence of the cosmic string, and the fractional part of the ratio of the magnetic flux by the quantum one, a closed expression for the respective Green function is obtained. Adopting this result, we explicitly calculate the renormalized vacuum expectation value of the energy-momentum tensors, $\langle T^A_B\rangle_{Ren}$, and analyse this result in some limitting cases.
\vspace{1pc}
\end{abstract}
\maketitle
\section{Introduction}
Cosmic strings are linear topologically stable gravitational defects which appear in the framework of grand unified theories. These objects could be produced in very early Universe as a result of spontaneous breakdown of gauge symmetry. Although topological defects have been first analysed in four-dimensional spacetime, they have been considered in the context of braneworld. In this scenario the defects live in a $n-$dimensions submanifold embedded in a $(4+n)-$dimensional Universe. In this context, the cosmic sting has been considered with two additional extra dimensions \cite{Cohen,Ruth}. Vacuum polarization effects associated with scalar and fermionic fields in four-dimensional cosmic string space-time, have been analyzed by many authors. It has been shown that these effects depend on the parameter which codify the conical structure of the geometry, $\alpha$.  Moreover, considering the presence of magnetic flux along the cosmic strings' core, there appears an additional contributions to the vacuum polarization effect associated with charged fields. 

The investigation of quantum effects in corresponding braneworld models is of considerable phenomenological interest, both in particle physics and cosmology. Quantum effects provide natural alternative for stabilizing the radion fields in a braneworld. The corresponding vacuum energy gives contribution to both the brane and bulk cosmological constant and, hence, has to be taking into account in the self-consistent formulation of the braneworld dynamics. Recently the fluxes by gauge fields play an important role in higher dimensional models including braneworld scenario (see for example \cite{Dou}).

Motivated by these facts, we analyse the vacuum polarization effects associated with massive charged spin$-1/2$ field in a six-dimensional cosmic string space-time considering a magnetic flux running along the core of the string. Specifically we are interested to calculate the vacuum expectation value of the energy-momentum tensor. In order to do that we present in a closed form, the expression for the fermionic Green function for a specific case where $\gamma$, the fractional part of the ratio of the magnetic flux by the quantum one, is equal to $\frac{(1-\alpha)}2$, and $\alpha=\frac1q$, being $q$ an integer number. Although being a very special situation, the analysis of vacuum polarization effects in this circumstance may shed light on the qualitative behavior of these quantities for non-integer $q$.

\section {The spinor Green function}
\label{sec2}
This section is devoted to calculate the Feynman propagator associated with a massive charged fermionic field propagating in a six-dimensional cosmic string space-time. Adopting the coordinate system, $x^M=(t,r,\varphi,x,y,z)$, the respective line element is given by:
\begin{eqnarray}
\label{cs}
	ds^2=g_{MN}dx^Mdx^N=-dt^2+dr^2+\alpha^2r^2d\varphi^2+\sum_{i=3}^5(dx^i)^2 \ , 
\end{eqnarray}
with $r\geq 0$, $\varphi\in[0, \ 2\pi]$, and $t, \ x^i\in(-\infty, \ \infty)$. The parameter $\alpha$ is smaller than unity and codify the presence of the string. In the braneworld scenario the space-time given by (\ref{cs}) represents a conical two-dimensional space transverse to a flat $3-$brane. (The core of the string is on the $3-$brane.)

In order to develop the calculation of the spinor Green function we shall adopt the $8\times 8$ Dirac matrices $\Gamma^M$ given below, which can be constructed in terms of the $4\times 4$ ones \cite{B-D,Moha}:
\begin{eqnarray}
\label{gamma}
\Gamma^0=\left( 
\begin{array}{cc}
0&\gamma^0 \\
\gamma^0&0 
\end{array} \right) \ , \
\Gamma^r=\left( 
\begin{array}{cc}
0&{\hat{r}}\cdot{\vec\gamma} \\
{\hat{r}}\cdot{\vec\gamma}&0 
\end{array} \right) \ , \
\Gamma^\varphi=\frac1{\alpha r}\left( 
\begin{array}{cc}
0&{\hat{\varphi}}\cdot{\vec\gamma} \\
{\hat{\varphi}}\cdot{\vec\gamma}&0 
\end{array} \right) \ , \nonumber\\
\Gamma^{x}=\left( 
\begin{array}{cc}
0&\gamma^{(3)}\\
\gamma^{(3)}&0 
\end{array} \right) \ , \
\Gamma^{y}=\left( 
\begin{array}{cc}
0&i\gamma_5\\
i\gamma_5&0 
\end{array} \right) \ , \
\Gamma^{z}=\left( 
\begin{array}{cc}
0&I \\
-I&0 
\end{array} \right) \ ,
\end{eqnarray}
where  $\gamma_5=i\gamma^0\gamma^1\gamma^2\gamma^3$, $I$ represents the $4\times 4$ identity matrix and ${\hat{r}}$ and ${\hat{\varphi}}$ stand the ordinary unit vectors in cylindrical coordinates. This set of matrices satisfies the Clifford algebra $\{\Gamma^M, \ \Gamma^N  \}=-2g^{MN}I_{(8)}$.

In the analysis of the vacuum polarization effects, we also consider the presence of an extra magnetic field running along the string. This magnetic field configuration is given by the following six-vector potential
\begin{eqnarray}
	A_M=A\partial_M\varphi
\end{eqnarray}
being $A=\frac{\Phi}{2\pi}$. 

The spinor Feynman propagator, defined as \cite{Birrel}
\begin{eqnarray}
	i{\cal{S}}_F(x,x')=\langle0|T(\Psi(x){\bar{\Psi}}(x'))|0\rangle \ ,
\end{eqnarray}
with ${\bar{\Psi}}=\Psi^\dagger\Gamma^0$, satisfies the non-homogeneous linear differential equation,
\begin{equation}
\label{Feyn}
\left(i\n +e\A -M\right){\cal{S}}_F(x,x')=\frac1{\sqrt{-g}}\delta^6(x-x')I_{(8)} \ ,
\end{equation}
where $g=det(g_{MN})$. The covariant derivative operator reads
\begin{equation}
\n=\Gamma^M(\partial_M+\Pi_M) \ ,
\end{equation}
being $\Pi_M$ the the spin connection given in terms of the $\Gamma-$matrices by
\begin{eqnarray}
\Pi_M=-\frac14\Gamma_N\nabla_M\Gamma^N \ ,
\end{eqnarray}
and
\begin{eqnarray}
\A=\Gamma^MA_M \ .
\end{eqnarray}

In \cite{Spin} we have shown that if a bispinor, ${\cal{D}}_F(x,x')$, satisfies the differential equation
\begin{eqnarray}
\label{D}
\left[{\Box}-ieg^{MN}(D_M A_N)+ie\Sigma^{MN}F_{MN}-2ieg^{MN}A_M\nabla_N\right. \nonumber\\
\left.-e^2g^{MN}A_M A_N-M^2-\frac14{\cal R}\right]{\cal{D}}_F(x,x')&=&-\frac1{\sqrt{-g}}\delta^6(x-x')I_{(8)} \ ,
\end{eqnarray}
with
\begin{equation}
\Sigma^{MN}=\frac14[\Gamma^M,\Gamma^N] \ , \ D_M=\nabla_M-ieA_M \ , 
\end{equation}
$\cal{R}$ being the scalar curvature and the generalized d'Alembertian operator 
given by
\begin{eqnarray}
\Box=g^{MN}\nabla_M\nabla_N=g^{MN}\left(\partial_M\nabla_N+\Pi_M\nabla_N-\{^S_{MN}\}\nabla_S\right) \ ,\nonumber
\end{eqnarray}
then the spinor Feynman propagator may be written as
\begin{equation}
\label{Sf} 
{\cal{S}}_F(x,x')=\left(i\n+e\A+M\right){\cal{D}}_F(x,x') \ .
\end{equation}

Applying this formalism for the system under investigation the operator $\cal{K}$, which acts on the left hand side of (\ref{D}), reads
\begin{eqnarray}
{\cal{K}}&=&{\Delta}+\frac i{\alpha^2 r^2}(1-\alpha)\Sigma^3_{(8)}\partial_\varphi-\frac1{4\alpha^2 r^2}(1-\alpha)^2+\frac e{\alpha^2 r^2} (1-\alpha)A\Sigma^3_{(8)}\nonumber\\
&-&\frac{2ie}{\alpha^2 r^2}A\partial_\varphi-\frac{e^2}{\alpha^2 r^2}A^2-M^2\ , 
\end{eqnarray}
where
\begin{equation}
\Sigma^3_{(8)}=\left( \begin{array}{cccc}
  \Sigma^3& 0\\ 
0 & \Sigma^3
                      \end{array}
               \right) \ , \ {\rm with} \
\Sigma^3=\left( \begin{array}{cccc}
  \sigma^3& 0\\ 
0 & \sigma^3
                      \end{array}
               \right) \ 
\end{equation}
and 
\begin{equation}
\label{delta}
{{\Delta}}=-\partial_t^2+\partial_r^2+\frac 1r\partial_r+\frac 1{\alpha^2 r^2}\partial^2_\varphi+\partial_x^2+\partial_y^2+\partial_z^2 \ .
\end{equation}

Due to the fact of $\Sigma^3_{(8)}$ be a diagonal matrix, the bispinor ${\cal{D}}_F(x,x')$ is diagonal, too. In this way we can obtain an expression for this Green function analyzing only the effective $2\times 2$ matrix differential equation below:
\begin{eqnarray}
\label{K}
\left[{\Delta}+\frac i{\alpha^2 r^2}(1-\alpha)\sigma^3\partial_\varphi-\frac1{4\alpha^2 r^2}(1-\alpha)^2+\frac e{\alpha^2 r^2} (1-\alpha)A\sigma^3\right.\nonumber\\
\left.-\frac{2ie}{\alpha^2 r^2}A\partial_\varphi-\frac{e^2}{\alpha^2 r^2}A^2-M^2\right]{\cal{D}}^{(2)}_F(x,x')= -\frac1{\sqrt{-g}}\delta^6(x-x')I_{(2)} \ .
\end{eqnarray}
So the complete Green function is given in terms of ${\cal{D}}^{(2)}_F(x,x')$, in a diagonal matrix form.

On basis of these results, for this six-dimensional cosmic string space-time, the fermionic propagator reads:
\begin{eqnarray}
	S_F(x,x')=\left[i\Gamma^0\partial_t+i\Gamma^r\partial_r+i\Gamma^\varphi\partial_\varphi+i\Gamma^i\partial_i-\frac{1-\alpha}2\Gamma^\varphi\Sigma^3_{(8)}+\frac {e\Phi}{2\pi}\Gamma^\varphi+M\right]D_F(x,x') \ .
\end{eqnarray}
                     
The vacuum average value for the energy-momentum tensor can be expressed in terms of the Euclidean Green function. It is related with the ordinary Feynman Green function \cite{Birrel} by the relation  ${\cal D}_E(\tau,\vec{r}; \tau', \vec{r'}) = -i {\cal D}_F(x,x')$, where $t=i\tau$. In the following we shall consider the Euclidean Green function. 

In order to obtain the Euclidean Green function ${\cal{D}}^{(2)}_E(x,x')$ in explicit form, let us find the complete set of bispinor which obey the eigenvalue equation
\begin{eqnarray}
\label{K1}
	{\bar{\cal{K}}}\Phi_\lambda(x)=-\lambda^2\Phi_\lambda(x)
\end{eqnarray}
with $\lambda^2\geq 0$ and being ${\bar{\cal{K}}}$ the Euclidean version of the differential operator given in (\ref{K}). So we may write
\begin{eqnarray}
\label{D2}
	{\cal{D}}^{(2)}_E(x,x')=\sum_{\lambda^2}\frac{\Phi_\lambda(x)\Phi_\lambda^\dagger(x')}{\lambda^2}=\int_0^\infty \ ds\sum_{\lambda^2} \Phi_\lambda(x)\Phi_\lambda^\dagger(x') \ e^{-s\lambda^2} \ .
\end{eqnarray}
The eigenfunctions of (\ref{K1}) can be specified by a set of quantum number associated with operators that commute with ${\bar{\cal{K}}}$ and among themselves: $p_\tau=-i\partial_\tau$, $p_i=-i\partial_i$ for $i=3,\ 4, 5$ , $L_\varphi=-i\partial_\varphi$ and the spin operator, $\sigma_3$. Let us denote these quantum numbers by $(k^\tau, \ k^i, \ n, \ \sigma)$, where $(k^\tau, \ k^i)\in (-\infty, \ \infty)$, $n=0, \ \pm 1, \ \pm 2, \ ... \ $, $\sigma=\pm 1$. Moreover, these functions also depend on the number $p$, which satisfies the relation $\lambda^2=p^2+k^2+M^2$ and assumes values in the interval $[0, \ \infty)$. On basis on these arguments the eigenfunction of the operator (\ref{K1}) is given by:

\begin{eqnarray}
\label{Phi}
	\Phi_\lambda^{(+)}(x)&=&\frac{e^{ik.x}{\sqrt{p}}}{[\alpha(2\pi)^{5}]^{1/2}}e^{in\varphi}J_{|\nu^+|/\alpha}(pr){\cal{\omega}}^{(+)} \ ,  \nonumber\\
	\Phi_\lambda^{(-)}(x)&=&\frac{e^{ik.x}{\sqrt{p}}}{[\alpha(2\pi)^{5}]^{1/2}}e^{in\varphi}J_{|\nu^-|/\alpha}(pr){\cal{\omega}}^{(-)}
\end{eqnarray}
where
	\begin{eqnarray}
	\label{omega}
	{\cal{\omega}}^{(+)}=\left(
\begin{array}{cc}
1 \\
0
\end{array} \right) \  , \ 
	{\cal{\omega}}^{(-)}=\left(
\begin{array}{cc}
0 \\
1
\end{array} \right) \ 
\end{eqnarray}
are the eigenfunctions of the the operator $\sigma_3$. $J_\mu(z)$ represents the Bessel functions of order
\begin{eqnarray}
\nu^{\pm}=n\pm\frac{(1-\alpha)}2-({\bar{N}}+\gamma)	\ .
\end{eqnarray}
We have defined $eA=e\frac{\Phi}{2\pi}={\bar{N}}+\gamma$, the ratio of the magnetic flux by the quantum one, in terms of an integer number, ${\bar{N}}$, and its fractional part, $\gamma$. 

Now we are in position to calculate the Green function by using (\ref{D2}), (\ref{Phi}) and (\ref{omega}) as show below
\begin{eqnarray}
	\label{D2a}
		{\cal{D}}^{(2)}_E(x,x')&=&\frac1{\alpha(2\pi)^{5}}\int_0^\infty ds  \int d^Nk \ \int_0^\infty \ dp \ p  \ e^{ik(x-x')}\sum_n e^{in(\varphi-\varphi')}\nonumber\\
	&& {\rm diag} (J_{|\nu^+|/\alpha}(pr)J_{|\nu^+|/\alpha}(pr'), J_{|\nu^-|/\alpha}(pr)J_{|\nu^-|/\alpha}(pr')) \ e^{-s(p^2+k^2+M^2)} \ .
\end{eqnarray}
With the help of \cite{Grad} we can express (\ref{D2a}) by
\begin{eqnarray}
\label{D2b}
	{\cal{D}}^{(2)}_E(x,x')&=&\frac1{\alpha(4\pi)^{3}}\int_0^\infty \frac{ds}{s^{3}}e^{-\frac{(\Delta x)^2+r^2+r'^2}{4s}-M^2s} \ \sum_n e^{in(\varphi-\varphi')}\nonumber\\
	&&{\rm diag}(I_{|\nu^+|/\alpha}(rr'/2s), \ I_{|\nu^-|/\alpha}(rr'/2s)) \ ,
\end{eqnarray}
where $I_\mu(z)$ is the modified Bessel function.

Unfortunately it is not possible to provide a closed result for the above integral in general case. However for a very special case it is. This will be the objective of the next two subsections.

\subsection{Massless case}
Taking $M=0$ in (\ref{D2b}), it is possible to integrate on the variable $s$ \cite{Grad}, and after some minor steps we find
\begin{eqnarray}
\label{D3d}
	{\cal{D}}_E^{(2)}(x',x)&=&\frac{e^{i{\bar{N}(\varphi-\varphi')}}}{16\pi^3\alpha(rr')^2\sinh^3u}\times\nonumber\\
&&	\left(\begin{array}{cc}
\cosh u S^{(+)}(u)-\sinh u S'^{(+)}(u)&0\\
0&\cosh u S^{(-)}(u)-\sinh S'^{(-)}(u)
\end{array} \right) \ , \nonumber\\ 
\end{eqnarray}
where
\begin{eqnarray}
\label{S}
	S^{(\pm)}(u)&=&\frac{e^{\mp i(\varphi-\varphi')}\sinh(\delta^\pm u/\alpha)-\sinh[(\delta^\pm-1)u/\alpha]}{\cosh(u/\alpha)-\cos(\varphi-\varphi')} \nonumber\\
	u&=&{\rm arccosh}{\left(\frac{(\Delta x)^2+r^2+r'^2}{2rr'}\right)} \ ,
\end{eqnarray}
being $\delta^\pm=\frac{(1-\alpha)}2\mp\gamma$.
In (\ref{D3d}) the prime denotes derivative with respect to $u$.

\subsection{Special case where $\alpha=\frac1q$ being $q$ an integer number}
\label{Heat0}
In this subsection we shall provide a closed expression for (\ref{D2b}) in the massive case. This will be done for 
\begin{eqnarray}
	\alpha=\frac1q \ \ {\rm and} \ \gamma=\frac{(1-\alpha)}2 \ ,
\end{eqnarray}
being $q$ an integer number. Although being a very special situation, the analysis of vacuum polarization effects in this circumstance may shed light on the qualitative behavior of these quantities for non-integer $q$. 

\subsubsection{Heat kernel}
\label{Heat}

Let us now investigate the effective $2\times 2$ diagonal matrix heat kernel, given in the integrand of (\ref{D2}). Using (\ref{D2b}), it is
\begin{eqnarray}
\label{Ka}
	{\cal{K}}(x,x';s)=\frac{e^{-\frac{(\Delta x)^2+r^2+r'^2}{4s}-M^2s}}{\alpha(4\pi s)^{3}}\sum_n e^{in\Delta\varphi}
	\left(\begin{array}{cc}
I_{|\nu^+|/\alpha}(rr'/2s)&0\\
0&I_{|\nu^-|/\alpha}(rr'/2s)
\end{array} \right) \ .
\end{eqnarray}

By using the formula \cite{Pru,Spin}, 
\begin{eqnarray}
	\sum_{k=0}^{q-1}e^{x\cos(\varphi+2\pi k/q)}=q\sum_{m=-\infty}^\infty I_{mq}(x)e^{imq\varphi} \ ,
\end{eqnarray}
it is possible to obtain closed expressions for the summations on the Bessel functions in (\ref{Ka}), and the equation (\ref{D2b}) becomes
\begin{eqnarray}
\label{Dq}
		{\cal{D}}_E^{(2)}(x',x)=\frac{e^{i{\bar{N}}\Delta\varphi}}{(2\pi)^{3}}M^{2}\sum_{k=0}^{q-1}\frac1{(\rho_k)^{2}}K_{2}(M\rho_k)	\left(\begin{array}{cc}
1&0\\
0&e^{i(1-1/q)\Delta\varphi}e^{-2i\pi k/q}
\end{array} \right) \ ,
\end{eqnarray}
with 
\begin{eqnarray}
\label{rho}
	\rho^2_k=(\Delta x)^2+r^2+r'^2-2rr'\cos\left(\Delta\varphi/q+2\pi k/q\right) \ ,
\end{eqnarray}
being $K_\mu$ the modified Bessel function. 

\section{Vacuum expectation value of the energy-momentum tensor}
\label{sec3}
The renormalized vacuum expectation value (VEV) of the energy-momentum tensor, $\langle T^A_B\rangle_{Ren.}$, must obey the conservation condition
	\begin{eqnarray}
\label{CC}
	\nabla_A\langle T^A_B \rangle_{Ren.}=0 \ ,
\end{eqnarray}
and for the physical situation under consideration having its trace being given by
\begin{eqnarray}
\langle T^A_A \rangle_{Ren.}=M^2{\rm Tr} \ {\cal{D}}_{Ren.}(x,x)	\ .
\end{eqnarray}

Using the point-splitting procedure \cite{Birrel}, the VEV of the energy-momentum tensor has the following form:
\begin{eqnarray}
\label{EM}
	\langle T_{AB}(x)\rangle=\frac 14\lim_{x'\to x}{ \rm Tr}\left[{\Gamma}_A(D_B-{\bar D}_{B'})+{\Gamma}_B(D_A-{\bar D}_{A'})\right]S_F(x,x') \ ,
\end{eqnarray}
where $D_M=\nabla_M-ieA_M$, and the bar denotes complex conjugate. Because the dependence of the fermionic Green function on the time variables, the zero-zero component of the energy-momentum tensor reads:
\begin{eqnarray}
\langle T_{00}(x)\rangle=\lim_{x'\to x}{\rm Tr} \ \Gamma_0\partial_0 S_F(x,x') \ , 
\end{eqnarray}
which can be expressed by 
\begin{eqnarray}
\label{T00}
	\langle T_{00}(x)\rangle=-i\lim_{x'\to x}\partial_t^2 \ {\rm Tr}\ {\cal{D}}_F(x',x)=-\lim_{x'\to x}\partial_\tau^2 \ {\rm Tr} \ {\cal{D}}_E(x,x') \ .	
\end{eqnarray}
In the obtainment of the above expression we have to use the fact that the bispinor ${\cal{D}}_F(x',x)$ is diagonal and ${\rm Tr} \ \Gamma^0\Gamma^i {\cal{D}}_E(x,x')=0$. 

Before to start the calculation let us analyse the bispinor (\ref{Dq}) in the coincidence limit. Taking $x'\to x$ we verify that it is divergent and that the divergence comes exclusively from the $k=0$ component. So, in order to obtain a finite and well defined result we should apply some renormalization prescription. This procedure can be applied in a manifest form by subtracting from (\ref{Dq}) its $k=0$ component. So the renormalized Green function is given by
\begin{eqnarray}
\label{Dren}
		{\cal{D}}_{Ren.}^{(2)}(x',x)=\frac{e^{i{\bar{N}}\Delta\varphi}}{(2\pi)^{3}}M^{2}\sum_{k=1}^{q-1}\frac1{(\rho_k)^{2}}K_{2}(M\rho_k)	\left(\begin{array}{cc}
1&0\\
0&e^{i(1-1/q)\Delta\varphi}e^{-2i\pi k/q}
\end{array} \right) \ .
\end{eqnarray}

Now let us proceed the calculation of $\langle T^M_N(x)\rangle_{Ren.}$. By using previous result,
\begin{eqnarray}
\label{T0m}
	\langle T_0^0(x)\rangle_{Ren.}=\lim_{x'\to x}\partial_\tau^2 \ {\rm Tr} \ {\cal{D}}_{Ren.}(x,x') 	\ ,
\end{eqnarray}
and substituting the Euclidean bispinor expressed by its diagonal form given in terms of (\ref{Dren}), after some intermediate steps we obtain 
\begin{eqnarray}
\label{T1m}
\langle T_0^0(x)\rangle_{Ren}=-\frac{M^{3}}{8\pi^3r^{3}}\sum_{k=1}^{q-1}\frac{\cos^2(k\pi/q)}{(\sin(k\pi/q) )^{3}}K_{3}(2Mr\sin(\pi k/q)) \ .
\end{eqnarray}
From (\ref{T1m}) we can observe that the energy density is non-positive quantity everywhere, vanishing only for $q=2$. For this case the contributions from the upper and lower component of the bispinor (\ref{Dren}) cancel each other. Moreover in the limit $Mr>>1/\sin(k\pi/q)$ the leading order is
\begin{eqnarray}
\langle T_0^0(x)\rangle_{Ren}\approx-\frac1{\pi(4\pi)^{3/2}\ r}\left(\frac Mr\right)^{5/2}\times\frac{\cos^2(\pi/q)}{(\sin(\pi/q))^{7/2}} e^{-2Mr\sin(\pi/q)}	\ ,
\end{eqnarray}
with exponentially suppressed behavior.

In the massless limit case, the energy density became expressed in terms of elementary functions,
\begin{eqnarray}
	\langle T_0^0(x)\rangle_{Ren}=-\frac1{8\pi^3r^6}\sum_{k=1}^{q-1}\frac{\cos^2(k\pi/q)}{(\sin(k\pi/q) )^{6}} \ .
\end{eqnarray}
By using the formulas
\begin{eqnarray}
\label{Ir}
	I_{N+2}(x)=\frac{I''_N(x)+N^2I_N(x)}{N(N+1)} \ {\rm and} \ I_2(x)=\frac{q^2}{\sin^2(qx)}-\frac1{\sin^2(x)} 
\end{eqnarray}
for the sum
\begin{eqnarray}
	I_N(x)=\sum_{k=1}^{q-1}\frac1{\sin^N(x+k\pi/q)} \ ,
\end{eqnarray}
we obtain
\begin{eqnarray}
\label{Tq}
	\langle T_0^0(x)\rangle_{Ren}=-\frac{(q^2-1)(q^4+q^2-20)}{3780\pi^3r^6} \ .
\end{eqnarray}
The above result is an analytical functions of $q$, and by analytical continuation is valid for all arbitrary values of $q$. Moreover, the above quantity is positive for $1<q \ < 2$. In figure $1$ is plotted the behavior of $\langle T_0^0\rangle_{Ren.}/M^{6}$ as function of $Mr$ for different values of the parameter $q$. We can see  that the modulus of the energy density increases when $q$ becomes larger.
\begin{figure}[tbph]
\label{figj2}
\begin{center}
\begin{tabular}{cc}
\epsfig{figure=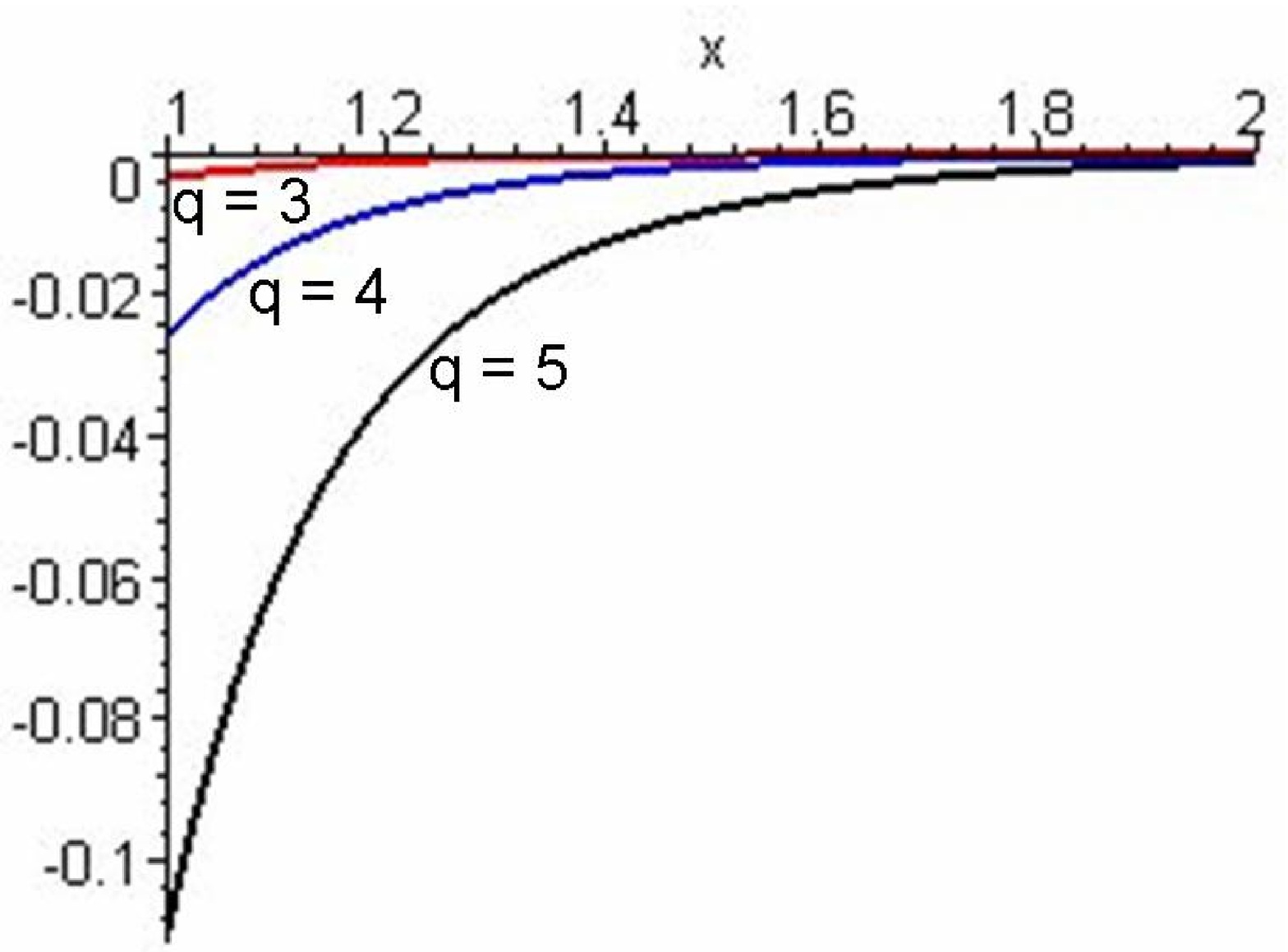, width=9.5cm, height=7.5cm,angle=0}
\end{tabular}
\end{center}
\end{figure}

Developing the calculation of the VEV of the energy density by using the bispinor (\ref{D3d}), the renormalized result is
\begin{eqnarray}
\label{Tq1}
\langle T_0^0(x)\rangle_{ren}&=-&\frac1{120960\pi^{3}\alpha^6  r^6}\left[-367(1-\alpha^2)^3+12(189\gamma^2+76)(1-\alpha^2)^2 \right. \nonumber \\
 && \left. +144(35\gamma^4-49\gamma^2-4)(1-\alpha^2)+1344\gamma^2(\gamma^2-1)(\gamma^2-4) \right] \ .
\end{eqnarray}
(Taking $\alpha=\frac1q$ and $\gamma=\frac{q-1}{2q}$ into the above expression we re obtain (\ref{Tq}).)

Now let us continue the calculation of the other components of the VEV of the energy-momentum tensor. Taking advantage of previous result we can promptly write down:
\begin{itemize}
\item The radial pressure, $\langle T^r_r\rangle$, can be calculated as shown below:
\begin{equation}
\langle T_{rr}(x)\rangle_{Ren.}=\frac{1}{2}\lim_{x'\to x}Tr\left[\Gamma_{r}\left(\partial_{r}-\partial_{r'} \right)\right]S_{F}\left( x,x'\right) \ .
\end{equation}
By using (\ref{Sf}) to calculate the  Feynman propagator, after a long calculation we find
\begin{eqnarray}
\label{Trr}
\langle {T_{r}^{r}}(x)\rangle_{Ren.}=-\frac{M^{3}}{8\pi^{3}r^{3}}\sum_{k=1}^{q-1}\frac{\cos^2(k\pi/q)}{\sin^{3}(k\pi/q)}K_{3}[2Mr\sin(k\pi/q)] \ .
\end{eqnarray}
\item From the conservation condition $\nabla_A\langle T^A_r\rangle_{Ren.}=0$, the azimuthal pressure, $\langle T^\varphi_\varphi\rangle$, can be obtained in terms of the radial one
\begin{eqnarray}
	\langle T^\varphi_\varphi(x)\rangle_{Ren.}=\partial_r\left(r\langle T^r_r(x)\rangle_{Ren.}\right) \ .
\end{eqnarray}
Substituting (\ref{Trr}) into the above expression we get
\begin{eqnarray}
\langle T_\varphi^\varphi(x)\rangle_{Ren.}&=&\frac{M^3}{8\pi^3r^3}\sum_{k=1}^{q-1}\frac{\cos^2(k\pi/q)}{\sin^3(k\pi/q)}\left\{5K_3[2Mr\sin(k\pi/q)] \right. \nonumber\\
&+&\left.2K_{2}[2Mr\sin(k\pi/q)]Mr\sin(k\pi/q)\right\} \ .
\end{eqnarray}
\item For the pressures along the directions parallel to the string we have (no summation over $i$)
\begin{eqnarray}
	\langle T^i_i\rangle_{Ren.}=\langle T^0_0\rangle_{Ren.} \ , \ i=3, \ 4, \ 5 \ .
\end{eqnarray}
\item Now on basis on the results obtained, we can verify the correct trace of the energy-momentum tensor:
\begin{eqnarray}
	\langle T_A^A(x)\rangle_{Ren.}=\frac{M^{4}}{4\pi^{3}r^{2}}\sum_{k=1}^{q-1}\frac{\cos^2(k\pi/q)}{\sin^2(k\pi/q)}K_{2}\left[ 2Mr\sin(k\pi/q) \right]=M^{2}\hbox{Tr}D_{Ren.}(x,x) \ .
\end{eqnarray}
\end{itemize}

\section{Conclusions and Discussions}
\label{conc}
In this paper we have investigated the local one-loop quantum gravity effects associated with massive charged fermionic fields in a six-dimensional cosmic string spacetime. As the first steps in the evaluation of the renormalized VEV of the energy-momentum tensor, in Section \ref{sec2} we have explicitly  provided the Euclidean Green function in an integral form. 

In Section \ref{sec3}, we have calculated the VEV of the energy-momentum tensor. Considering $\alpha=\frac1q$, being $q$ an integer number, and $\gamma=\frac{q-1}{2q}$, the corresponding renormalized VEV of the energy-momentum tensor is expressed in a closed form for massive fields. We explicitly shown the behaviors of the energy-density for several limiting cases. Moreover, we calculated completely all components of the energy-momentum tensor, and shown they satisfy the trace identity.

An interesting point which deserves to be mentioned is that the effects on the renormalized VEV of the energy-momentum tensor due to the conical structure of the spacetime and the magnetic interaction, may cancel each other. This was explicitly observed in Section \ref{sec3} for $q=2$ and $\gamma=\frac14$, for the energy-densities given in (\ref{T1m}) and (\ref{Tq1}).

\end{document}